\documentclass[prb,showpacs,twocolumn,aps]{revtex4-1}
\usepackage[utf8]{inputenc}

\usepackage{graphicx,graphics}

\usepackage{latexsym}
\usepackage{amssymb}
\usepackage{amsfonts}
\usepackage{epsfig}
\usepackage{psfrag}
\usepackage{graphicx}

\usepackage{epsfig,color}
\usepackage{psfrag}
\usepackage[utf8]{inputenc} 
\definecolor{darkgreen}{rgb}{0,0.4,0}

\newcommand{\bea}{\begin{eqnarray}}
\newcommand{\ea}{\end{eqnarray}}
\newcommand{\eea}{\end{eqnarray}}

\usepackage{amsmath}

\usepackage{color}
\renewcommand{\ll}{\langle\langle}

\begin{document}

%\title{Loughborough Group. Report LG1. 2020.06.08}
\title{Towards the Heisenberg limit in microwave photon detection by a qubit array}
\author{P. Navez, A. G. Balanov, S. E. Savel'ev, A. M. Zagoskin}
\begin{abstract}
%Single photon detection could prove to be a new tool to detect the missing dark matter in the universe. 
%We present a scheme for such a device in a quantum circuit consisting in a qubit array interacting with a waveguide.
Using an analytically solvable model, we show that a qubit array-based detector allows to achieve the fundamental Heisenberg limit in detecting single photons. In case of superconducting qubits, this opens new opportunities for quantum sensing and communications in the important microwave range. 
%is investigated in order to assess its performance. In the case of correlated qubit state the Heisenberg limit for the signal-to-noise ratio is achieved. We estimate the signal to noise ratio for various entanglement arrangments of the qubit states and show that high connectivity between the qubits is necessary for an efficient detection.
\end{abstract}

\affiliation{
Department of Physics, Loughborough University,
Loughborough LE11 3TU, United Kingdom
}

\date{\today}

\maketitle

%\section{KIT hamiltonian}
%\begin{widetext}

\section{Introduction}
Single photon detection has been initiated in the optical regime, but is now being actively pursued in the microwave range. Its motivations come from different fields, ranging from quantum communications\cite{ren2017ground}, sensing\cite{PhysRevX.9.021019} and medical imaging\cite{kiani2020quantum} to the search for axions, hypothetical ingredients of dark matter\cite{Duffy_2009,BECK20156,dixit2020searching} (an axion in a magnetic field would convert to a single microwave photon). 

The natural frequency range of superconducting qubits is in the microwave regime\cite{Zagoskin2011}, which facilitates their use for  detection of single microwave photons\cite{Schuster2007,Inomata2016,GU20171,PhysRevLett.112.093601}. Still, it remains a challenging task, in particular because of the significant level of ambient noise in this frequency range. 

%It was predicted in Ref.\cite{Zagoskin_2013} that a collective readout of an array of $N$ uncorrelated qubits would allow detection of an external signal, even at a single-photon level, in the presence of local uncorrelated noise. The signal-to-noise ratio (SNR) had the standard $\sqrt{N}$-dependence, corresponding to standard quantum limit (SQL). 
The fast progress in quality and complexity of experimentally realized superconducting  qubit arrays coupled to waveguides \cite{PhysRevLett.107.240501,PhysRevA.76.042319,wendin2017quantum,wallraff2004strong,PhysRevLett.91.097906,Hoi_2013} as well as in theory of these quantum metamaterials\cite{rakhmanov2008quantum,zagoskin2009quantum,savel2012two,ivic2016qubit,grimsmo2020quantum} stimulated the inquiry into the practical possibility of their application to single photon detection. It was shown\cite{Zagoskin_2013} that for an array of ${\cal N}$ uncorrelated, non-interacting qubits used for a quantum non-demolition detection of a photon, the signal-to-noise ratio (SNR) 
ratio grows as $\sqrt{\cal N}$, corresponding to standard standard quantum limit (SQL)
%{\color{red} shot noise limit}, 
%{\color{red} SHOT NOISE limit instead ???}, 
and it was suggested \cite{Giovannetti2006}that introducing some specific quantum correlations could realize the so-called Heisenberg limit for SNR, i.e., to have SNR scaling as ${\cal N}$ rather than $\sqrt{\cal{N}}$.

Essentially, SQL is not {\em the} fundamental limit on the accuracy achieved by ${\cal N}$ measurements of the system (in parallel or consecutively). It applies to the case of independent inputs and is the consequence of the central limit theorem\cite{Giovannetti2004}. If the inputs are properly quantum correlated, then the accuracy is proportional to $1/\cal{N}$ rather than $1/\sqrt{\cal{N}}$, and this is the best possible outcome\cite{Giovannetti2006}. Remarkably, achieving the Heisenberg limit does not depend on whether the {\em measurements} are quantum correlated; it is only the quantum correlations between the {\em inputs} that count\cite{Giovannetti2006}. 

Several possible implementations for achieving the Heisenberg limit had been suggested\cite{Giovannetti2006}, but its experimental realization remains a challenging task, especially outside the optical range. In this paper we show that the Heisenberg limit for SNR can be achieved in principle using quantum nondemolition measurement of an array of qubits, which plays the role of an antenna.
%we investigate in detail the possibility of achieving the Heisenberg limit in an array of qubits employed as a detector. 
%Using an analytically solvable model we conclude that this is indeed possible. 
This result may be of a particular importance for superconducting qubit arrays, which would be a natural choice for the detection of single microwave photons\cite{Zagoskin_2013}. For convenience and without the loss of generality, we are using the language appropriate to this particular realization of the proposed approach.
In section 2, we formulate an exactly solvable model Hamiltonian in presence of a thermal noise and, in section 3, we analyze the obtained results  according to the level of input correlations between the qubits. 

%The last progress on quantum circuit however have opened the window  for  investigating the range  of microwave radiation interacting with qubits under various configurations (transmon, phase and flux qubit, ...)  \cite{Zagoskin_2013,zagoskin2011quantum,zagoskin2009quantum,rakhmanov2008quantum,zagoskin2014test,PhysRevA.86.065803,PhysRevA.100.023844}. The improvement 
%on longer coherence times has rendered possible extremely sensitive detection in the microwave domain. 
%The constant development in superconducting circuits has been extremely intensive over the last decades in order to reduce considerably the external noise 
%associated to any electronic component in particular to the Josephson junctions \cite{PhysRevLett.107.240501,PhysRevA.76.042319,wendin2017quantum}. 

\section{Exactly solvable model}

\subsection{Model Hamiltonian}

 Without loss of generality, we start from the scheme presented in Fig.1. There are $N$ identical waveguides indexed by $\nu=1,2,\dots N$, each containing $M$ identical qubits and $\mu=1,2,\dots M$. These qubits are prepared  either in a factorized superposition state $|\Psi_F\rangle = \prod_{\nu,\mu}|\psi\rangle_{\nu,\mu} = \prod(|0\rangle_{\nu,\mu} +|1 \rangle_{\nu,\mu})/\sqrt{2}$, or in a more complicated entangled state chosen to maximize the SNR.

An input signal consists of $n$ photons ($n=0,1$) and is uniformly fed into the same mode of waveguides. As in \cite{Zagoskin_2013}, the qubits are detuned from the mode frequency (dispersive regime) to ensure that there is no absorption. Thus the qubit-field interaction is a quantum non-demolition kind of process.

\begin{figure}
\begin{center}
 \includegraphics[width=8cm]{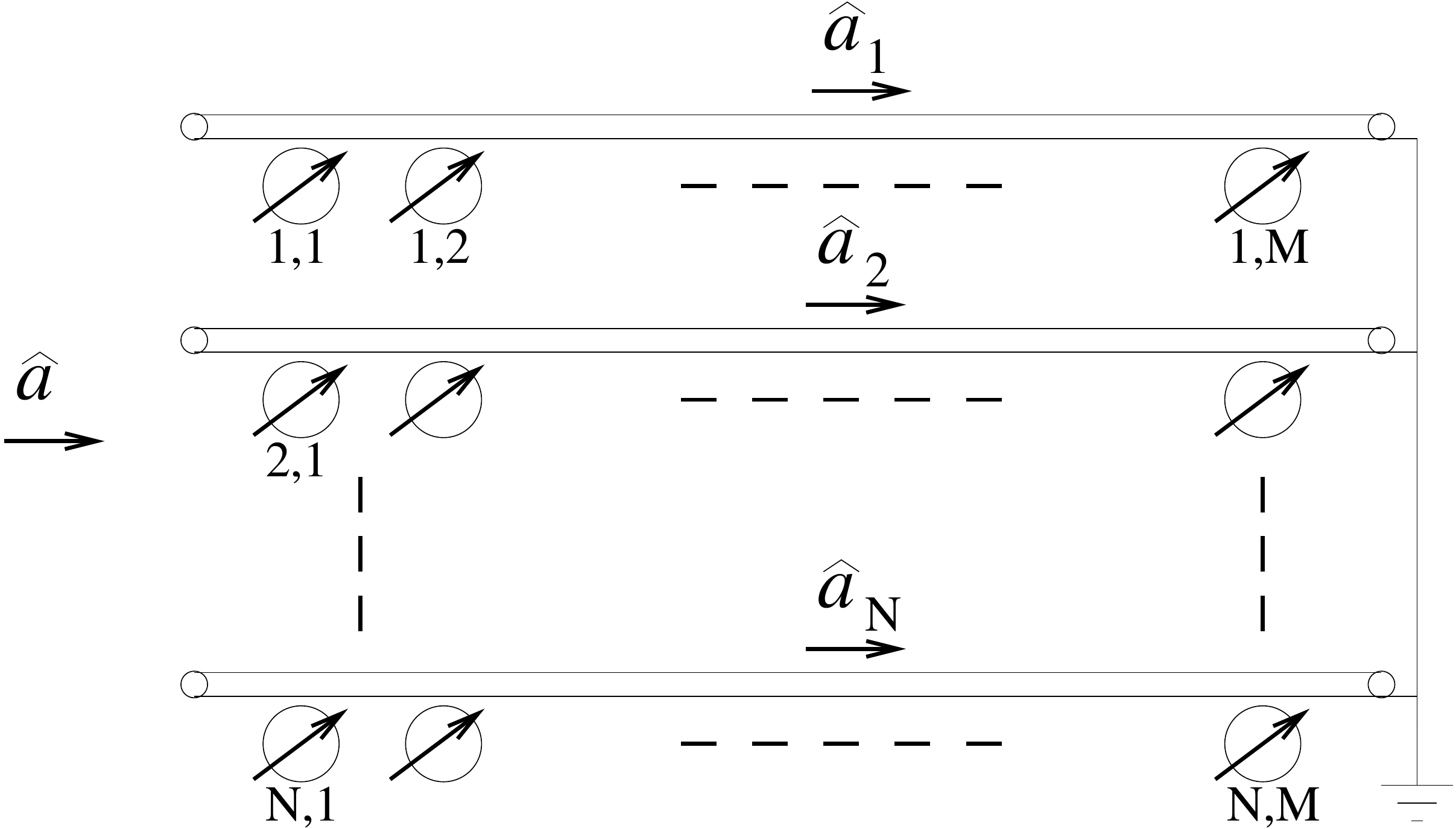}
 % qbitarray.eps: 2954x1429 px, 300dpi, 25.01x12.10 cm, bb=0 0 709 343
\end{center}
\caption{Schematic representation of the single photon detector. The signal propagates  horizontally
%along the $z$-axis 
through $N$ parallel waveguides each containing $M$ qubits.}
\end{figure}

The annihilation operator for the input signal, $\hat a$, can be then written as
\begin{eqnarray}
\hat a=\sum_{\nu=1}^{N} 
%\frac{e^{-i2\pi k_x x/N}}
\frac{1}{\sqrt{N}}\hat a_{\nu} 
\end{eqnarray}
Here $\hat a_\nu$ describes the submode, which travels in the waveguide $\nu$. The qubit is described by Pauli matrices $\hat\sigma_{\nu,\mu}^{x}=|1\rangle_{\nu,\mu} \langle 0| + |0 \rangle_{\nu,\mu} \langle 1|$, $\hat\sigma_{\nu,\mu}^{y}=-i|1\rangle_{\nu,\mu} \langle 0| + i|0 \rangle_{\nu,\mu} \langle 1|$ and $\hat \sigma_{\nu,\mu}^z=|1 \rangle_{\nu,\mu} \langle 1| - |0 \rangle_{\nu,\mu} \langle 0|$. The rising/lowering operators are $\hat \sigma_{\nu,\mu}^\pm=(\hat \sigma_{\nu,\mu}^x \pm i \hat \sigma_{\nu,\mu}^y)/2$. 
%A transmon qubit, however, has higher excited states as well,  $|n \rangle_{\nu,\mu}$. 
%If we constrain the system's evolution to a Hilbert subspace spanned by $|0\rangle_{\nu,\mu}$ and   $|1\rangle_{\nu,\mu}$,  these operators are reduced to Bose creation/annihilation operators via $\hat \sigma_{\nu,\mu}^z\simeq 2\hat c_{\nu,\mu}^\dagger \hat c_{\nu,\mu}-1$, $\hat \sigma_{\nu,\mu}^- \simeq \hat c_{\nu,\mu}$ and $\hat \sigma_{\nu,\mu}^+ \simeq \hat c_{\nu,\mu}^\dagger$. This is the key approximation, which reduces the problem to an exactly solvable one. {\color{red} Not clear! That does not make the problem exactly solvable! The use of creation/annihilation operators  is for the case of transmons in case of higher occupancy not in the regime of anharmonicity and should be distinguished from qubit.}
%In as such with a polarization initially in the x direction, the qubits do not interact with the waveguide fields. A preliminary small coupling given by the Hamiltonian term $h_\delta \sigma^x_{\nu,\mu}$ during  a interval time $t=\tau_1$ changes the polarization and exposes the qubits to a subsequent photo detection. It results in a tilted rotation of the qubit about the x axis with an angle $\Delta =\tau_1 h_\delta \ll 1$. 

Even in the absence of direct qubit-qubit interaction, they will interact through the virtual excitations of the vacuum waveguide mode, which in case of a large number of qubits can be approximated by the mean field term in the Hamiltonian proportional to $h_\delta \hat \sigma^x_{\nu,\mu}$ \cite{Zagoskin_2013}. During the relevant time interval $\tau_1$ this interaction induces a rotation of the qubit state about $0x$ by an angle $\Delta =\tau_1 h_\delta \ll 1$.
The interaction of qubits with the photon field in the dispersive regime is given by a term $\propto h_p\hat a^\dagger_\nu \hat a_\nu \hat \sigma^z_{\nu,\mu}$. When one photon passes by a qubit (with an effective interaction time $\tau_2$), it will induce a qubit state rotation about the  $0z$ axis by $\theta= \tau_2 h_p$. If the distance  between the qubits is much smaller than the photon wavelength, 
we can assume that the interaction with all qubits is simultaneous.
%The whole photon radiation passage lasts for the total time $\tau_2+ L/c \simeq \tau_2$ where the photon passage time $L/c$ through the waveguide of size  $L$ is assumed to not contribute much.  
In the absence of noise, the time-dependent  Hamiltonian can be thus written as
\begin{eqnarray}
\hat H(t)&=& -\bigg[\sum_{\nu,\mu=1}^{N,M} R \left(\frac{t}{\tau_1}-\frac{1}{2}\right) 
\frac{h_\delta}{2}\hat \sigma^x_{\nu,\mu} 
\nonumber \\
&+&
R\left(\frac{t-\tau_1}{\tau_2}-\frac{1}{2}\right) \frac{h_p}{2}\hat \sigma^z_{\nu,\mu} 
\hat a^\dagger_\nu \hat a_\nu \bigg]
\end{eqnarray}
where we use the rectangular window function:
\begin{eqnarray}
 R(t)=\left\{{\begin{array}{rl}0,&{\text{if }}|t|>{\frac {1}{2}}\\{\frac {1}{2}},&{\text{if }}|t|={\frac {1}{2}}\\1,&{\text{if }}|t|<{\frac {1}{2}}.\end{array}}\right.
\end{eqnarray}

%The interaction of the photon field is a quantum non demolition process obtained by shifting the energy level of the two qubit states given by the Hamiltonian term $h_p \sigma^z_{\nu,\mu}$. When one photon passes through one qubit with a duration time $\tau_2$, a second tilted rotation about the  z axis is achieved with an angle $\theta= \tau_2 h_p$. The whole photon radiation passage lasts for the total time $\tau_2+ L/c \simeq \tau_2$ where the photon passage time $L/c$ through the waveguide of size  $L$ is assumed to not contribute much.  
\subsection{Noise}

We consider simultaneous action of  kinds of noise. The first kind is inherent in the quantum state chosen to describe the qubit ensemble. The readout measurement has an intrinsic shot noise from uncorrelated qubits that is probed from the quantum fluctuations of the spin operators. We expect that an adequate quantum entanglement would correlate the qubit states and decrease this measurement noise.

The second kind of noise originates from the imperfect transmission of the signal to the qubits and leads to the dephasing. It may be caused by imperfections of the waveguide, thermal photons inside the cavity or in the qubits, or other sources of ambient noise. It is modeled by the additional quantum noise field $\hat b_{\nu,\mu}$ that reduces the quantum efficiency $\eta <1$. As a result the actual signal interacting with the qubit $(\nu,\mu)$ will be
\begin{eqnarray}\label{noise}
\hat a'_{\nu,\mu}=\sqrt{\eta}\hat a_{\nu} + \sqrt{1- \eta} \hat b_{\nu,\mu}
\end{eqnarray}
For a thermal noise field, the disturbance will depend on the effective noise temperature $T$. Note that this effect accounts also for the back scattering  field:  
\begin{eqnarray}\label{noise'}
\hat b'_{\nu,\mu}=-\sqrt{1- \eta}\hat a_{\nu} + \sqrt{\eta} \hat b_{\nu,\mu} \,.
\end{eqnarray}
%these correspond to the orthogonal contributions that do not interact with the qubits.} 

We express the noise through its Fourier components $\hat b_{\nu,k_z}$ inside each waveguide: 
\begin{eqnarray}
\hat b_{\nu,\mu}= \sum_{\mu=1}^M \frac{e^{2\pi ik_z \mu /M}}{\sqrt{M}}\hat b_{\nu,k_z}
\end{eqnarray}
These evolves with the dispersion relation frequency $\omega_{k_z}$. Therefore
\begin{eqnarray}
\hat a'_{\nu,k_z}=\sqrt{\eta} \hat a_{\nu} + \sqrt{1-\eta} \hat b_{\nu,k_z}
\end{eqnarray}
With all these contributions, the initial quantum state has many degrees of freedom. Modeling the noise in the waveguide by a black body radiation,  
%of the waveguide described in the canonical  thermal ensemble, 
we have for the initial density matrix of the system the following expression: 
\begin{eqnarray}
\!\!\!\!\hat \rho_n= \frac{\exp(-\sum_{\nu,k_z=1}^{N,M} \frac{\hbar \omega_{k_z}}{{k_B T}} \hat b^\dagger_{\nu,k_z} \hat b_{\nu,k_z})}
{{\rm Tr} [\exp(
-\sum_{\nu,k_z=1}^{N,M} \frac{\hbar \omega_{k_z}}{{k_B T}} \hat b^\dagger_{\nu,k_z} \hat b_{\nu,k_z})]} |n\rangle \langle n| |\Psi \rangle \langle \Psi |.  
\end{eqnarray}
Here $|n\rangle = (1/\sqrt{n!})[\frac{\sum_{\nu=1}^{N}\hat a^\dagger_{\nu}}{\sqrt{N}}]^n|0\rangle$ is the state of the signal photon field with $n(=0, 1)$ photons, and $|\Psi\rangle$ is the quantum state of the qubits. 

The other sources of noise are the relaxation of the qubits and the effects of finite preparation and readout timescales. We have described by a single effective decay rate $\Gamma$.
%Another source of noise is the relaxation of the qubits with a decay rate $\Gamma$.
%along the $z$ direction that occurs when they behave independantly. 
It affects the initial state 
%prepared in the $x$ direction 
since the qubit readout is delayed by the finite time interval $\tau_1 +\tau_2$ during the photon passage.
%with a decay rate $\Gamma$. 
If no photon passes through the line, the quantum state becomes mixed with the density matrix:
\begin{eqnarray}
|\psi\rangle_{\nu,\mu}\langle \psi | \rightarrow 
\hat \rho_{\nu,\mu} (t)=
\nonumber \\
\frac{1}{2} 
\left[\exp(-\Gamma t)\hat \sigma_{\nu,\mu}^x 
+(\exp(-2\Gamma t)-1)\hat \sigma_{\nu,\mu}^z + 
\hat 1 \right]
%\langle \hat \sigma_{\nu,\mu}^z(\tau_1 + \tau_2) \rangle=
%\exp(-\tau_2/2\Gamma)(1+ \langle \hat \sigma_{\nu,\mu}^z(\tau_1)\rangle)-1 
\end{eqnarray}
Therefore, we must ensure that the photon passage time should be short enough, so that $\tau_1+ \tau_2 \ll 1/\Gamma$ (see also \cite{PhysRevA.97.022115,Danilin_2018}). This condition can be fullfilled since the decay rate is in the MHz range in comparison to the photon frequency in the GHz range. %Another way to prevent this relaxation is the introduction of interaction between the qubits. 
The addition of direct qubit-qubit 
%An adequate 
coupling 
generates an entangled ground state, stabilizes the initial state $|\Psi_F \rangle$ and suppresses the relaxation rate (see section \ref{3B}).

\subsection{System evolution}

In the Heisenberg representation, the two successive qubit rotations (by an angle $\Delta$ during the time $\tau_1$ due to effective qubit-qubit interactions, and by an angle 
$\theta$ during the time $\tau_2$ due to direct qubit-photon interaction) produce a unitary transformation of field operators. For the qubit, we obtain after the total time $\tau_1+\tau_2$ the new ``spin'' operators:
\begin{eqnarray}
\hat \sigma'^i_{\nu,\mu} &=&\hat U_\sigma 
\hat \sigma_{\nu,\mu}^i  \hat U^\dagger_\sigma 
\quad i=(x,y,z)  
\\
\hat U_\sigma &= &\prod_{\nu,\mu=1}^{N,M}\exp(-i\frac{\theta}{2} \hat a'^\dagger_{\nu,\mu}\hat a'_{\nu,\mu}\hat \sigma^z_{\nu,\mu}) \exp(-i\frac{\Delta}{2} \hat \sigma^x_{\nu,\mu})
\end{eqnarray}

%The qubit lowering operator evolves according to
%\begin{eqnarray}
%&&\hat c'_{\nu,\mu} =\hat U 
%\hat c_{\nu,\mu} \hat U^\dagger  
%\\ 
%\hat U &=&
%\prod_{\nu,\mu=1}^{N,M}
%\exp(-i\theta \hat c^\dagger_{\nu,\mu} \hat c_{\nu,\mu}  \hat a'^\dagger_{\nu,\mu}\hat a'_{\nu,\mu})\exp(-i\frac{\Delta}{2} (\hat c_{\nu,\mu}^\dagger + \hat c_{\nu,\mu}))\, . 
%\nonumber \\
%\end{eqnarray}
Using the operator algebra, we bring these expressions to a simpler form,
\begin{eqnarray}\label{s1}
\hat \sigma'^z_{\nu,\mu} &=&\cos(\Delta)
\hat \sigma_{\nu,\mu}^z  -\sin(\Delta)\cos(\theta \hat a'^\dagger_{\nu,\mu}\hat a'_{\nu,\mu})
\hat \sigma_{\nu,\mu}^y 
\nonumber \\
&&+\sin(\Delta)\sin(\theta \hat a'^\dagger_{\nu,\mu}\hat a'_{\nu,\mu})
\hat \sigma_{\nu,\mu}^x 
%\\ \label{t1}
%\hat c'_{\nu,\mu} &=&\exp(i\theta \hat a'^\dagger_{\nu,\mu}\hat a'_{\nu,\mu})
%\hat c_{\nu,\mu} +i\Delta/2   
\end{eqnarray}
It remains to determine the expectation values of these operators for various qubit configurations using the definition 
$\langle \hat{\cal O} \rangle= {\rm tr}(\hat \rho_n \hat {\cal O})$.

\section{Results}

\subsection{Uncorrelated initial qubit state}

The simplest qubit configuration is the  factorized state given by
$|\Psi_F \rangle = \prod_{\nu,\mu=1}^{N,M}
(|0\rangle_{\nu,\mu}+ |1 \rangle_{\nu,\mu})/\sqrt{2}$. 
Using the Eqs.(\ref{s1})
%,\ref{t1}) 
and observing that the signal affects equally each qubit, we obtain for the average magnetic flux of the array:
\begin{eqnarray}
\langle \hat S'^z\rangle&=&\!\! \!\sum_{\nu,\mu=1}^{N,M}\!\!\!\frac{\langle \hat \sigma'^z_{\nu,\mu} \rangle}{2}=\frac{NM}{2}\sin(\Delta)\langle \sin(\theta \hat a'^\dagger_{\nu,\mu}\hat a'_{\nu,\mu})\rangle\,.
\end{eqnarray}
%and for the average number of excitation:
%\begin{eqnarray}
%\langle \hat N'_t\rangle&=&\sum_{\nu,\mu=1}^{N,M}
%\langle \hat  c'^\dagger_{\nu,\mu} \hat c'_{\nu,\mu}  \rangle 
%\nonumber\\
%&=&\frac{NM}{2}[\Delta \langle\sin(\theta \hat a'^\dagger_{\nu,\mu}\hat a'_{\nu,\mu}) \rangle +\Delta^2/2 +1]  
%\end{eqnarray}
Quite generally, we obtain for the detection of $n$ photons (see Appendix):
\begin{widetext}
\begin{eqnarray}
\langle \sin(\theta \hat a'^\dagger_{\nu,\mu}\hat a'_{\nu,\mu})\rangle_n= 
\frac{-i/2}{1+f_T(1-e^{i\theta})}
\left[1+\frac{\eta(e^{i\theta}-1)}{N[1+f_T(1-e^{i\theta})}\right]^n   + {\rm c.c}
\end{eqnarray}
where $f_T=(1-\eta)\overline{f}$ is the fraction of thermal noise deteriorating the signal, $\overline{f}=\sum_{k_z=1}^M f_{k_z}/M$ is the noise density and $f_{k_z}=1/[\exp(\hbar\omega_{k_z}/k_B T)-1]$ is the Bose-Einstein factor.
In particular, we find for $n=0,1$
\begin{eqnarray}
\langle \sin(\theta \hat a'^\dagger_{\nu,\mu}\hat a'_{\nu,\mu})\rangle_0= \frac{f_T \sin\theta}{
[1 + f_T(1-\cos\theta)]^2+f_T^2 \sin^2\theta}
\end{eqnarray}
\begin{eqnarray}
\langle \sin(\theta \hat a'^\dagger_{\nu,\mu}\hat a'_{\nu,\mu})\rangle_1=
\langle \sin(\theta \hat a'^\dagger_{\nu,\mu}\hat a'_{\nu,\mu})\rangle_0+
\frac{\eta\sin\theta(1- 2(1-\cos\theta) f_T^2)}
{N\{[1 + f_T(1-\cos\theta)]^2+f_T^2 \sin^2\theta\}^2}
\end{eqnarray}
\end{widetext}
For small $\Delta$, we deduce also that the fluctuations scale normally like $NM$ since 
\begin{eqnarray}
\langle \delta^2 \hat S'^z\rangle=
NM/4+ {\cal O}(\Delta) 
%\\
%\langle \delta \hat N'^2_t\rangle=
%\frac{NM}{4}+ {\cal O}(\Delta)
\end{eqnarray}
Thus, the  SNR for an $n$-photon detection is defined as
\begin{eqnarray}\label{s2}
SNR_n =\frac{\langle \hat S'^z \rangle}
{\sqrt{\langle \delta^2  \hat S'^z\rangle}}
%=\frac{\langle \hat N'_t \rangle}
%{\sqrt{\langle \delta^2  \hat N'_t\rangle}}
=R \sqrt{N} \Delta\langle \sin(\theta \hat a'^\dagger_{\nu,\mu}\hat a'_{\nu,\mu})\rangle_n
\nonumber \\
\end{eqnarray}
where $R=\sqrt{M}$ corresponds to the standard quantum limit.
This last result implies that, as a function of $N$, SNR scales like $\sqrt{N}$ for small $\Delta$ in the absence of a photon  but becomes worse in the presence of one photon. However, as a function of $M$, the scaling is always  $\sqrt{M}$ irrelevant of the number of photons detected. This observation indicates that $N=1$ is the optimal value for a single photon detection. 
%{\color{red} It is indeed impossible to generate multiple copies of a single quantum state due to the no-cloning theorem.}
%Essentially, this is the result of Ref.\cite{Zagoskin_2013}.

The contrast is defined as:
\begin{eqnarray}
{\cal C}\equiv \frac{\langle \sin(\theta \hat a'^\dagger_{\nu,\mu}\hat a'_{\nu,\mu})\rangle_1- \langle \sin(\theta \hat a'^\dagger_{\nu,\mu}\hat a'_{\nu,\mu}) \rangle_0 }
{\langle \sin(\theta \hat a'^\dagger_{\nu,\mu}\hat a'_{\nu,\mu})\rangle_0+\langle \sin(\theta \hat a'^\dagger_{\nu,\mu}\hat a'_{\nu,\mu}) \rangle_1} 
\end{eqnarray}
This expression can be reduced to
\begin{eqnarray}
\frac{N}{\eta}\frac{2{\cal C}}{1-{\cal C}}=
\frac{1- 2(1-\cos\theta) f_T^2}
{f_T[1 + 2(1-\cos\theta)(f_T +f_T^2)] }
\end{eqnarray}
The left hand side corresponds to a renormalized contrast 
where the dependence from the quantum efficiency and the qubit number $N$ have been absorbed 
and its curve is displayed in Fig.2. As expected, the contrast decreases when the noise density increases, but this effect is more important for large rotation angles $\theta$. It shows a regime of negative contrast when the noise density exceeds the critical value $f_T > 1/\sqrt{2(1-\cos\theta)}.$ For rather small noise, $f_T<0.5$, the contrast is positive for any $\theta\in(0,\pi)$. With this, $C>0$ for any  finite $f_T$ if qubit-photon interaction does not cause much qubit rotation for vanishing $\theta$.
% especially for rotation angles close to $\pi$. 

\begin{figure}
 \centering
\includegraphics[width=9cm]{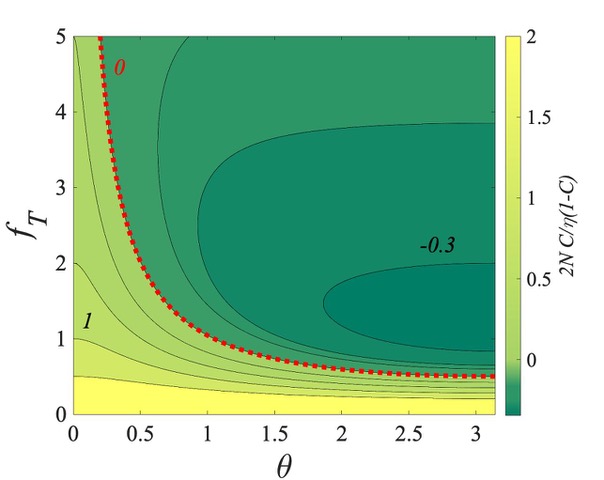}
 % contrast1.pdf: 0x0 px, 300dpi, 0.00x0.00 cm, bb=
 \caption{Plot of the renormalized contrast vs. the angle shift and the thermal noise}
 \label{fig:2}
\end{figure}

\subsection{Perfectly correlated state} \label{3B}

%Since the signal with one photon scales like 
%$1/N$, there is no interest to correlate the states with different $N$. 
Correlated states involving qubits with equal $\nu$ but different $\mu$ improve the signal-to-noise ratio. More precisely, using the notation $\hat S_{\nu}^i=\sum_{\mu=1}^{M}
\hat \sigma_{\nu,\mu}^i/2$, we notice  from (\ref{s2}) that  SNR  for small $\Delta$ is proportional to $R= \langle \hat S_{\nu}^x \rangle/\langle \hat {S_{\nu}^z}^2 \rangle^{1/2}$ which  equals to $\sqrt{M}$ in the uncorrelated case.
This observation suggests that the optimal state is the one that maximizes $R$ and therefore is the ground state of  the quantum Ising Hamiltonian with a strong qubit-qubit coupling:
%along the $z$ axis for a fixed $x$:
\begin{eqnarray}
\hat H_\nu= 
\hat S_{\nu}^{z\,2}  +\lambda 
\hat S_{\nu}^x 
\end{eqnarray}
The coefficient $\lambda$ is the Lagrange multiplier that controls the necessary level of correlation. It should be small since  an optimal result is obtained for strong quantum correlations. 
It is worth   pointing out this connection between the  magnetometry quantum sensing in cold atoms and 
the single photon detection using qubit detection. The optimization of SNR amounts to obtaining the highest possible spin squeezing \cite{Cirac}.
Then we may use the perturbation theory. In the zeroth order ($\lambda \rightarrow 0$, we start from the total angular momentum state  $|l,m \rangle_\nu$ as the eigenstate of ${\bf \hat S}^2_{\nu} = \sum_{i=x,y,z}  \hat S_{\nu}^{i\,2}$ and $\hat S_{\nu}^z$ with  eigenvalues $l(l+1)$  and $m$ respectively and where we identify $l=M/2$. The ground state combines states of $m$ very close to zero and thus corresponds  to a giant spin state that entangles all the individual qubits. 

In the next order, there are two cases:

\medskip

{\bf $M$ odd:}
The next order is a linear combination of the two degenerate ground states 
 $|M/2,\pm 1/2 \rangle_\nu$. We find the  correlated state $|\Psi \rangle=\prod_{\nu=1}^{N}
(|M/2,1/2\rangle_{\nu}+ |M/2,-1/2 \rangle_{\nu})/\sqrt{2}$ that corresponds to a maximally entangled state.
We determine that 
$\langle \delta^2 \hat S^z_\nu\rangle=1/4$ is independent on $M$ and  that
$\langle \hat S^x_\nu\rangle= (M+1)/4$. We find therefore  $R=(M+1)/2$ which corresponds asymptotically to the Heisenberg limit.

\medskip

{\bf $M$ even:}
The ground state for $\lambda=0$ corresponds to a maximally 
entangled state of individual qubits:
\begin{eqnarray}
|M/2, 0\rangle_\nu=\frac{(M/2)!}{\sqrt{M!}}
\!\!\!\!\!\!
\sum_{\begin{array}{c} \tiny
s_\mu=\pm 1/2   \\   \tiny                   
\!\!\!\!\!\!\!  \sum_{\mu=1}^M s_\mu=0
   \end{array}}
\!\!\!\!\!\!\!|s_1,\dots,s_\mu,\dots,s_M \rangle_\nu
\end{eqnarray}
with spin $s_\mu$ along the z axis.
Each individual state is also an eigenstate for 
$\lambda=0$.

The expansion to the second order involves the entangled combination $|\Psi \rangle=\prod_{\nu=1}^{N}
\sum_{m=-1}^1 \alpha_m |M/2, m\rangle_{\nu}$. 
We determine 
$\langle \delta^2 \hat S^z_\nu\rangle=\lambda^2(M/2+1)M/4$ and 
$\langle \hat S^x_\nu\rangle=\lambda(M/2+1)M/2$. We find therefore  $R=\sqrt{M(M/2+1)}$ which also asymptotically tend to the Heisenberg limit which agrees with the results for squeezed spin states  in \cite{PhysRevA.65.053819}.

\begin{figure}
 \centering
 \includegraphics[width=9cm]{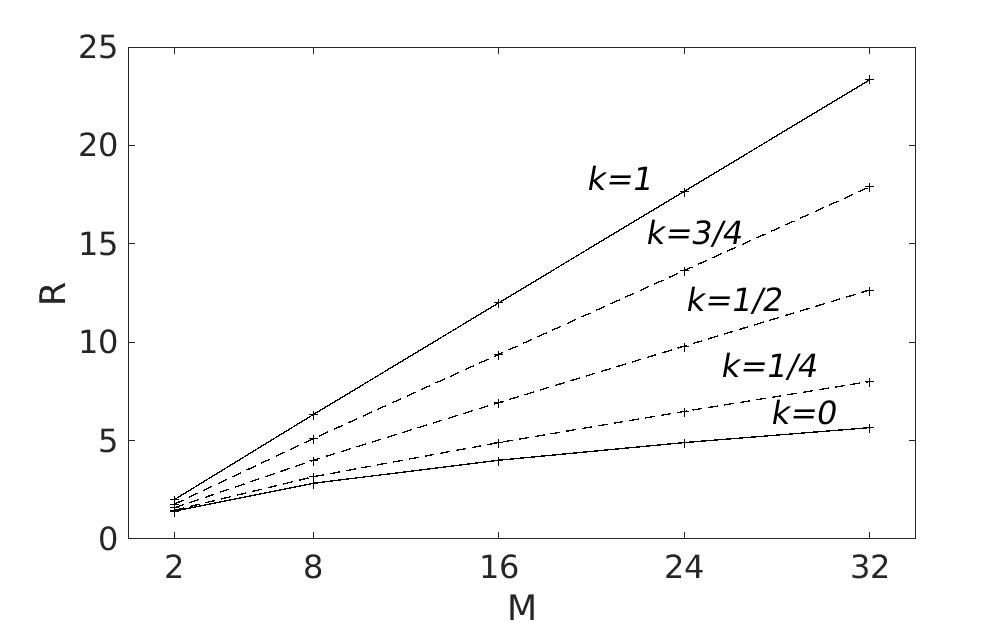}
 % Heisl.jpg: 580x390 px, 72dpi, 20.46x13.76 cm, bb=0 0 580 390
 \caption{Signal to noise ratio $R$ vs. qubit number $M$ for perfect correlated state ($k=1$), non correlated state ($k=0$) and intermediate regimes ($0<k<1$)}
 \label{fig:3}
\end{figure}

Such a maximally correlated state of the qubit set, which imposes the maximal rigidity on it producing a ``giant spin", is not realistic from the experimental point of view. Generally, restrictions imposed by the qubit circuit geometry limit their connectivity or couplings to architecture in clusters. 
%However,  
A relaxation of effective qubit-qubit coupling  
%does not quickly destroy 
affects the rigidity and deteriorate the SNR. Indeed, the introduction of a pertubation additional small term (which softens the "giant spin") 
\begin{eqnarray}
\hat H'_\nu=
%\epsilon 
\sum_{\mu\not= \mu'=1} ^{M}p_{\mu \mu'}\frac{\hat \sigma^z_{\nu,\mu} \hat \sigma^z_{\nu,\mu'}}{4} 
%\quad \quad \quad \epsilon \ll 1
\end{eqnarray}
raises the degeneracy of the individual states. Assume a simplified case where the interaction is turned on between $M-M_E$ spins and does not affect the  $M_E$ other spins. For $M_E$ even, the state that minimizes the perturbation Hamiltonian for $\lambda=0$ can be decomposed into an uncorrelated state of 
$M-M_E$ spins and a perfectly correlated state of $M_E$ spins described by angular state with angular momentum number $M_E/2$ written as: 
\begin{eqnarray}
|\Psi\rangle_\nu= 
%\frac{1}{\sqrt{2}}\sum_\pm
|\pm s_1,\dots,\pm s_{M-M_E}\rangle  |M_E/2, 0\rangle_{\nu}
\end{eqnarray}
In this expression, there are two possible degenerate states of opposite uncorrelated spins. 
%We took the sum combination over these two as the minimum  as thehe sum  
Again, as the result of second order perturbation theory, we obtain the expression for $R$,
\begin{eqnarray}
R&=&\sqrt{M_E(M_E/2+1)+M-M_E}
\end{eqnarray}
which interpolates  between the uncorrelated regime and the Heisenberg regime. Fig. \ref{fig:3} shows the increase of the SNR with the qubit number $M$ for various entanglement state   characterized by the ratio $k=M_E/M$. 
This last result illustrates that the Heisenberg limit of SNR is not fragile with respect to perturbations.

\medskip

{\bf $M=2$  for any $\lambda$:}
For this simpler situation, the exact diagonalization is possible. Using the correlated state  $|\Psi \rangle=\prod_{\nu=1}^{N}
\sum_{m=-1}^1 \alpha_m |M/2, m\rangle_{\nu}$, we determine the coefficient $\alpha_m$ and find the ratio:
\begin{eqnarray}
R=\sqrt{2+ \frac{2}{\sqrt{1+4\lambda^2}}} 
\end{eqnarray}
Thus $\lambda$ controls the crossover from 
an uncorrelated state ($R=\sqrt{2}$ at $\lambda=\infty$) to a fully correlated or entangled state ($R=2$ at $\lambda=0$), with the corresponding transition from the SQL to the Heisenberg limit.

\subsection{Nearest-neighbour interacting state}

Since the qubits are placed in a linear chain along the waveguide, we can use   the quantum Ising Hamiltonian with only nearest-neighbour coupling:
%along the $z$ axis for a fixed $x$:
\begin{eqnarray}
\hat H_\nu=  \sum_{\mu=1}^{M}\frac{\hat \sigma^z_{\nu,\mu+1} \hat \sigma^z_{\nu,\mu} }{2} +\lambda\frac{\hat \sigma^x_{\nu,\mu} }{2}
\end{eqnarray}
where we use periodic boundary condition for simplicity ($\hat \sigma^z_{\nu,M+1}=\hat \sigma^z_{\nu,1}$).

We define the antiferromagnetic state 
$|AF \rangle_\nu$ as the ground state for $\lambda=0$. These are two such degenerate states but we will only consider one since they are separated by a large potential barrier for large $N$.  In the  case of $\lambda\to 0$, we use again the second order  perturbation theory with the correlated state $|\Psi \rangle=\prod_{\nu=1}^{N}
( \alpha_0 + \alpha_1 \hat S^+_\nu + \alpha_{-1} \hat S^-_\nu)|AF\rangle_{\nu}$. 
There are again two cases:

\medskip

{\bf $M$ even:}
We find:
$\langle \delta^2\hat S^z_\nu\rangle=\lambda^2 M/16$ and 
$\langle \hat S^x_\nu\rangle=\lambda M/4$. Therefore we find no improvement since  $R=\sqrt{M}$ is left unchanged.

%which is the standard noise limit. 

%Probably, improvement can be made around the critical point (transition paramagnetic-antiferromagnetic). 

\medskip

{\bf $M$ odd:}
We find for large $M$ that 
$\langle \delta^2 \hat S^z_\nu\rangle=(1+\lambda^2 M/4)/4$ and $\langle  \hat S^x_\nu\rangle=\lambda M/4$. We find therefore  $R=\lambda M/\sqrt{4+\lambda^2 M} \rightarrow \sqrt{M}$ for large $M$.
In the limit of small $\lambda$ however, we find $R=\lambda M/2$ which is again the  Heisenberg limit, but only achieved for  $\lambda \ll 1$.

Therefore the nearest-neighbour coupling is useless if the goal is to reach the Heisenberg limit of signal-to-noise ratio in a quantum detector array.

\section{Conclusion}

We have analyzed a detector consisting in an array of qubits in the presence of noise and interaction. Using an exactly solvable model, we have shown that quantum correlations of the state of array allow to reach the Heisenberg limit, that is, increase the signal-to-noise ratio proportionally to the total number of qubits (forming together the detected signal) rather than its square root. The rigidity of the qubit system is the crucial requirement for achieving this effect, which makes connectivity between qubits in a detector array the key parameter in designing similar systems.
Our findings provide an important insight for strategy of  reaching the Heisenberg limit, in the design of efficient qubit-based quantum detectors.

In our generic model we have neglected the size of the qubits compared to the microwave wavelength. We also do not specify the origin of the interaction, which produces the interqubit coupling. A natural extension of this study will be the investigation of different geometries of qubit arrays and of the effects of specifics of the interqubit interaction (dipole-dipole, inductive, etc.).

%{\color{red} It is also worth noticing that parity in the qubit number $M$ influences significatively the SNR for a perfectly correlated state by a greater prefactor for even number. }

%We have considered an analytically solvable model for a 2D qubit array sensor. We demonstrate the importance of using an entangled state of qubits in order to increase the signal to noise ratio, and show that the Heisenberg limit can be achieved. Limitations arise from the practical implementation of full entanglement between the qubits, in particular, the difficulty associated to a full 
%connectivity between them. The case of nearest neighbour interaction produces a correlated antiferromagnetic state that generally does not improve the signal to noise ratio. 

{\bf Acknowledgements:}
The authors thank K. Shulga for helpful discussions. 
This work was supported by the EU project SUPERGALAX
(Grant agreement ID: 863313). 
%a grant of the Russian Science Foundation (Project No. 17-12-01587).

\begin{widetext}

\section{Appendix}

%\subsection{Extension to many modes (model 1.1)}
The expectation value of the phase is determined using 
the coherent state formalism.
Defining the coherent state: 
$|\alpha \rangle_{\nu} =\sum_{n=0}^\infty e^{-|\alpha|^2/2}\alpha^n (\hat a^\dagger_{\nu})^n |0\rangle_{\nu,\mu} /n!$ and $|\beta \rangle_{\nu,\mu} =\sum_{n=0}^\infty e^{-|\beta|^2/2}\beta^n (\hat b^\dagger_{\nu,\mu})^n |0\rangle_{\nu,\mu}/n!$, we can calculate successively the coherent state average:
\begin{eqnarray}
_{\nu} \langle \alpha|_{\nu,\mu}\langle \beta | 
\exp[i\theta \hat a'^\dagger_{\nu,\mu} \hat a'_{\nu,\mu}]
|\alpha \rangle_{\nu}|\beta \rangle_{\nu,\mu}&=&
_{\nu} \langle \alpha|_{\nu,\mu}\langle \beta | 
\exp[i\theta (\sqrt{\eta} \hat a^\dagger_{\nu} + \sqrt{1-\eta} \hat b^\dagger_{\nu,\mu})
(\sqrt{\eta} \hat a_{\nu} + \sqrt{1-\eta} \hat b_{\nu,\mu})]
|\alpha \rangle_{\nu}|\beta \rangle_{\nu,\mu}
\nonumber \\
&=&_{\nu} \langle 0|_{\nu,\mu}\langle 0 | 
\exp[i\theta (\hat a'^\dagger_{\nu,\mu} +\sqrt{\eta} \alpha^* + \sqrt{1-\eta} \beta^*)
(\hat a'_{\nu,\mu} +\sqrt{\eta} \alpha + \sqrt{1-\eta} \beta)]
|0 \rangle_{\nu}|0 \rangle_{\nu,\mu}
\nonumber \\
&=&_{\nu,\mu} \langle \sqrt{\eta} \alpha + \sqrt{1-\eta} \beta  | 
\exp[i\theta \hat a'^\dagger_{\nu,\mu} 
\hat a'_{\nu,\mu} ]
|\sqrt{\eta} \alpha + \sqrt{1-\eta} \beta \rangle_{\nu,\mu}
\nonumber \\
&=&
_{\nu,\mu} \langle \sqrt{\eta} \alpha + \sqrt{1-\eta} \beta  | 
e^{i\theta}(\sqrt{\eta} \alpha + \sqrt{1-\eta} \beta) \rangle_{\nu,\mu}
\nonumber \\
&=& 
\exp[(e^{i\theta}-1)|\sqrt{\eta} \alpha + \sqrt{1-\eta} \beta|^2]
\end{eqnarray}
The thermal state caused by the noise is related to the coherent state through the formula:
\begin{eqnarray}
\frac{\exp(-\hbar \omega_{k_z} \hat b^\dagger_{\nu,k_z} \hat b_{\nu,k_z}/k_B T)}
{{\rm Tr} [\exp(-\hbar \omega_{k_z} \hat b^\dagger_{\nu,k_z} \hat b_{\nu,k_z}/k_B T)]} 
=\int \frac{d^2\beta}{\pi f_{k_z}} \exp(-|\beta|^2/f_{k_z}) |\beta \rangle_{\nu,k_z} \langle \beta |
\end{eqnarray}
With this procedure, we obtain an expression for the thermal average:
\begin{eqnarray}
&&{\rm Tr} [|\alpha \rangle_{x} \langle \alpha|
\frac{\exp(-\sum_{k_z=1}^{M}\hbar \omega_{k_z} \hat b^\dagger_{\nu,k_z} \hat b_{\nu,k_z}/k_B T)}
{{\rm Tr} [\exp(-\sum_{k_z=1}^{M}\hbar \omega_{k_z} \hat b^\dagger_{\nu,k_z} \hat b_{\nu,k_z}/k_B T)]}
\exp[i\theta \hat a'^\dagger_{\nu,\mu} \hat a'_{\nu,\mu}]]
\nonumber\\
&=&\prod_{k_z=1}^{M} \int \frac{d^2\beta_{k_z}}{\pi f_{k_z}} \exp(-|\beta_{k_z}|^2/f_{k_z}) 
\exp[(e^{i\theta}-1)|\sqrt{\eta} \alpha + \sqrt{1-\eta} 
\sum_{k_z=1}^{M}\frac{\exp(i2\pi k_z \nu/M)}{\sqrt{M}}\beta_{k_z}|^2]
\nonumber \\
&=& 
\frac{1}{1+f_T(1-e^{i\theta})}
\exp\left[\frac{\eta(e^{i\theta}-1)|\alpha|^2}{1+f_T(1-e^{i\theta})}\right]
\end{eqnarray}
It remains to generalize the expressions for $N$ channels and 
carry out an expansion development in the photon number.
Using the property that $|\alpha \rangle= \prod_{\nu=1}^{N}|\alpha/\sqrt{N} \rangle_{\nu}$, we determine
\begin{eqnarray}
\langle \exp[i\theta  a'^\dagger_{\nu,\mu} \hat a'_{\nu,\mu}] \rangle&=& {\rm Tr} [|\alpha\rangle \langle \alpha|
\frac{\exp(-\sum_{\nu',k_z=1}^{N,M}\hbar \omega_{k_z} \hat b^\dagger_{\nu',k_z} \hat b_{\nu',k_z}/k_B T)}
{{\rm Tr} [\exp(-\sum_{\nu',k_z=1}^{N,M}\hbar \omega_{k_z} \hat b^\dagger_{\nu',k_z} \hat b_{\nu',k_z}/k_B T)]}
\exp[i\theta  a'^\dagger_{\nu,\mu} \hat a'_{\nu,\mu}]
\nonumber \\
&=&
\frac{1}{1+f_T(1-e^{i\theta})}
\exp\left[\frac{\eta(e^{i\theta}-1)|\alpha|^2}{N[1+f_T(1-e^{i\theta})}\right]
\end{eqnarray} 
Finally, by an expansion in $\alpha$, we deduce the average for the case of $n$ photons:
\begin{eqnarray}
\langle \exp[i\theta  a'^\dagger_{\nu,\mu} \hat a'_{\nu,\mu}] \rangle_n &=& {\rm Tr} [|n \rangle \langle n|
\frac{\exp(-\sum_{\nu', k_z=1}^{N,M}\hbar \omega_{k_z} \hat b^\dagger_{\nu',k_z} \hat b_{\nu',k_z}/k_B T)}
{{\rm Tr} [\exp(-\sum_{\nu',k_z=1}^{N,M}\hbar \omega_{k_z} \hat b^\dagger_{\nu',k_z} \hat b_{\nu',k_z}/k_B T)]}
\exp[i\theta  a'^\dagger_{\nu,\mu} \hat a'_{\nu,\mu}]
\nonumber \\
&=&
\frac{1}{1+f_T(1-e^{i\theta})}
\left[1+\frac{\eta(e^{i\theta}-1)}{N[1+f_T(1-e^{i\theta})}\right]^n
\end{eqnarray}

\end{widetext}

\bibliographystyle{apsrev4-1}
%\bibliography{DESEPTICON-PRB-1}
\bibliography{paperrefs2}

\begin{thebibliography}{29}%
\makeatletter
\providecommand \@ifxundefined [1]{%
 \@ifx{#1\undefined}
}%
\providecommand \@ifnum [1]{%
 \ifnum #1\expandafter \@firstoftwo
 \else \expandafter \@secondoftwo
 \fi
}%
\providecommand \@ifx [1]{%
 \ifx #1\expandafter \@firstoftwo
 \else \expandafter \@secondoftwo
 \fi
}%
\providecommand \natexlab [1]{#1}%
\providecommand \enquote  [1]{``#1''}%
\providecommand \bibnamefont  [1]{#1}%
\providecommand \bibfnamefont [1]{#1}%
\providecommand \citenamefont [1]{#1}%
\providecommand \href@noop [0]{\@secondoftwo}%
\providecommand \href [0]{\begingroup \@sanitize@url \@href}%
\providecommand \@href[1]{\@@startlink{#1}\@@href}%
\providecommand \@@href[1]{\endgroup#1\@@endlink}%
\providecommand \@sanitize@url [0]{\catcode `\\12\catcode `\$12\catcode
  `\&12\catcode `\#12\catcode `\^12\catcode `\_12\catcode `\%12\relax}%
\providecommand \@@startlink[1]{}%
\providecommand \@@endlink[0]{}%
\providecommand \url  [0]{\begingroup\@sanitize@url \@url }%
\providecommand \@url [1]{\endgroup\@href {#1}{\urlprefix }}%
\providecommand \urlprefix  [0]{URL }%
\providecommand \Eprint [0]{\href }%
\providecommand \doibase [0]{http://dx.doi.org/}%
\providecommand \selectlanguage [0]{\@gobble}%
\providecommand \bibinfo  [0]{\@secondoftwo}%
\providecommand \bibfield  [0]{\@secondoftwo}%
\providecommand \translation [1]{[#1]}%
\providecommand \BibitemOpen [0]{}%
\providecommand \bibitemStop [0]{}%
\providecommand \bibitemNoStop [0]{.\EOS\space}%
\providecommand \EOS [0]{\spacefactor3000\relax}%
\providecommand \BibitemShut  [1]{\csname bibitem#1\endcsname}%
\let\auto@bib@innerbib\@empty
%</preamble>
\bibitem [{\citenamefont {Ren}\ \emph {et~al.}(2017)\citenamefont {Ren},
  \citenamefont {Xu}, \citenamefont {Yong}, \citenamefont {Zhang},
  \citenamefont {Liao}, \citenamefont {Yin}, \citenamefont {Liu}, \citenamefont
  {Cai}, \citenamefont {Yang}, \citenamefont {Li} \emph
  {et~al.}}]{ren2017ground}%
  \BibitemOpen
  \bibfield  {author} {\bibinfo {author} {\bibfnamefont {J.-G.}\ \bibnamefont
  {Ren}}, \bibinfo {author} {\bibfnamefont {P.}~\bibnamefont {Xu}}, \bibinfo
  {author} {\bibfnamefont {H.-L.}\ \bibnamefont {Yong}}, \bibinfo {author}
  {\bibfnamefont {L.}~\bibnamefont {Zhang}}, \bibinfo {author} {\bibfnamefont
  {S.-K.}\ \bibnamefont {Liao}}, \bibinfo {author} {\bibfnamefont
  {J.}~\bibnamefont {Yin}}, \bibinfo {author} {\bibfnamefont {W.-Y.}\
  \bibnamefont {Liu}}, \bibinfo {author} {\bibfnamefont {W.-Q.}\ \bibnamefont
  {Cai}}, \bibinfo {author} {\bibfnamefont {M.}~\bibnamefont {Yang}}, \bibinfo
  {author} {\bibfnamefont {L.}~\bibnamefont {Li}},  \emph {et~al.},\
  }\href@noop {} {\bibfield  {journal} {\bibinfo  {journal} {Nature}\ }\textbf
  {\bibinfo {volume} {549}},\ \bibinfo {pages} {70} (\bibinfo {year}
  {2017})}\BibitemShut {NoStop}%
\bibitem [{\citenamefont {Santagati}\ \emph {et~al.}(2019)\citenamefont
  {Santagati}, \citenamefont {Gentile}, \citenamefont {Knauer}, \citenamefont
  {Schmitt}, \citenamefont {Paesani}, \citenamefont {Granade}, \citenamefont
  {Wiebe}, \citenamefont {Osterkamp}, \citenamefont {McGuinness}, \citenamefont
  {Wang}, \citenamefont {Thompson}, \citenamefont {Rarity}, \citenamefont
  {Jelezko},\ and\ \citenamefont {Laing}}]{PhysRevX.9.021019}%
  \BibitemOpen
  \bibfield  {author} {\bibinfo {author} {\bibfnamefont {R.}~\bibnamefont
  {Santagati}}, \bibinfo {author} {\bibfnamefont {A.~A.}\ \bibnamefont
  {Gentile}}, \bibinfo {author} {\bibfnamefont {S.}~\bibnamefont {Knauer}},
  \bibinfo {author} {\bibfnamefont {S.}~\bibnamefont {Schmitt}}, \bibinfo
  {author} {\bibfnamefont {S.}~\bibnamefont {Paesani}}, \bibinfo {author}
  {\bibfnamefont {C.}~\bibnamefont {Granade}}, \bibinfo {author} {\bibfnamefont
  {N.}~\bibnamefont {Wiebe}}, \bibinfo {author} {\bibfnamefont
  {C.}~\bibnamefont {Osterkamp}}, \bibinfo {author} {\bibfnamefont {L.~P.}\
  \bibnamefont {McGuinness}}, \bibinfo {author} {\bibfnamefont
  {J.}~\bibnamefont {Wang}}, \bibinfo {author} {\bibfnamefont {M.~G.}\
  \bibnamefont {Thompson}}, \bibinfo {author} {\bibfnamefont {J.~G.}\
  \bibnamefont {Rarity}}, \bibinfo {author} {\bibfnamefont {F.}~\bibnamefont
  {Jelezko}}, \ and\ \bibinfo {author} {\bibfnamefont {A.}~\bibnamefont
  {Laing}},\ }\href {\doibase 10.1103/PhysRevX.9.021019} {\bibfield  {journal}
  {\bibinfo  {journal} {Phys. Rev. X}\ }\textbf {\bibinfo {volume} {9}},\
  \bibinfo {pages} {021019} (\bibinfo {year} {2019})}\BibitemShut {NoStop}%
\bibitem [{\citenamefont {Kiani}\ \emph {et~al.}(2020)\citenamefont {Kiani},
  \citenamefont {Villanyi},\ and\ \citenamefont {Lloyd}}]{kiani2020quantum}%
  \BibitemOpen
  \bibfield  {author} {\bibinfo {author} {\bibfnamefont {B.~T.}\ \bibnamefont
  {Kiani}}, \bibinfo {author} {\bibfnamefont {A.}~\bibnamefont {Villanyi}}, \
  and\ \bibinfo {author} {\bibfnamefont {S.}~\bibnamefont {Lloyd}},\
  }\href@noop {} {\enquote {\bibinfo {title} {Quantum medical imaging
  algorithms},}\ } (\bibinfo {year} {2020}),\ \Eprint
  {http://arxiv.org/abs/2004.02036} {arXiv:2004.02036 [quant-ph]} \BibitemShut
  {NoStop}%
\bibitem [{\citenamefont {Duffy}\ and\ \citenamefont {van
  Bibber}(2009)}]{Duffy_2009}%
  \BibitemOpen
  \bibfield  {author} {\bibinfo {author} {\bibfnamefont {L.~D.}\ \bibnamefont
  {Duffy}}\ and\ \bibinfo {author} {\bibfnamefont {K.}~\bibnamefont {van
  Bibber}},\ }\href {\doibase 10.1088/1367-2630/11/10/105008} {\bibfield
  {journal} {\bibinfo  {journal} {New Journal of Physics}\ }\textbf {\bibinfo
  {volume} {11}},\ \bibinfo {pages} {105008} (\bibinfo {year}
  {2009})}\BibitemShut {NoStop}%
\bibitem [{\citenamefont {Beck}(2015)}]{BECK20156}%
  \BibitemOpen
  \bibfield  {author} {\bibinfo {author} {\bibfnamefont {C.}~\bibnamefont
  {Beck}},\ }\href {\doibase https://doi.org/10.1016/j.dark.2015.03.002}
  {\bibfield  {journal} {\bibinfo  {journal} {Physics of the Dark Universe}\
  }\textbf {\bibinfo {volume} {7-8}},\ \bibinfo {pages} {6 } (\bibinfo {year}
  {2015})}\BibitemShut {NoStop}%
\bibitem [{\citenamefont {Dixit}\ \emph {et~al.}(2020)\citenamefont {Dixit},
  \citenamefont {Chakram}, \citenamefont {He}, \citenamefont {Agrawal},
  \citenamefont {Naik}, \citenamefont {Schuster},\ and\ \citenamefont
  {Chou}}]{dixit2020searching}%
  \BibitemOpen
  \bibfield  {author} {\bibinfo {author} {\bibfnamefont {A.~V.}\ \bibnamefont
  {Dixit}}, \bibinfo {author} {\bibfnamefont {S.}~\bibnamefont {Chakram}},
  \bibinfo {author} {\bibfnamefont {K.}~\bibnamefont {He}}, \bibinfo {author}
  {\bibfnamefont {A.}~\bibnamefont {Agrawal}}, \bibinfo {author} {\bibfnamefont
  {R.~K.}\ \bibnamefont {Naik}}, \bibinfo {author} {\bibfnamefont {D.~I.}\
  \bibnamefont {Schuster}}, \ and\ \bibinfo {author} {\bibfnamefont
  {A.}~\bibnamefont {Chou}},\ }\href@noop {} {\enquote {\bibinfo {title}
  {Searching for dark matter with a superconducting qubit},}\ } (\bibinfo
  {year} {2020}),\ \Eprint {http://arxiv.org/abs/2008.12231} {arXiv:2008.12231
  [hep-ex]} \BibitemShut {NoStop}%
\bibitem [{\citenamefont {Zagoskin}(2011)}]{Zagoskin2011}%
  \BibitemOpen
  \bibfield  {author} {\bibinfo {author} {\bibfnamefont {A.~M.}\ \bibnamefont
  {Zagoskin}},\ }\href@noop {} {\emph {\bibinfo {title} {Quantum engineering:
  theory and design of quantum coherent structures}}}\ (\bibinfo  {publisher}
  {Cambridge University Press},\ \bibinfo {year} {2011})\BibitemShut {NoStop}%
\bibitem [{\citenamefont {Schuster}\ \emph {et~al.}(2007)\citenamefont
  {Schuster}, \citenamefont {Houck}, \citenamefont {Schreier}, \citenamefont
  {Wallraff}, \citenamefont {Gambetta}, \citenamefont {Blais}, \citenamefont
  {Frunzio}, \citenamefont {Majer}, \citenamefont {Johnson}, \citenamefont
  {Devoret}, \citenamefont {Girvin},\ and\ \citenamefont
  {Schoelkopf}}]{Schuster2007}%
  \BibitemOpen
  \bibfield  {author} {\bibinfo {author} {\bibfnamefont {D.~I.}\ \bibnamefont
  {Schuster}}, \bibinfo {author} {\bibfnamefont {A.~A.}\ \bibnamefont {Houck}},
  \bibinfo {author} {\bibfnamefont {J.~A.}\ \bibnamefont {Schreier}}, \bibinfo
  {author} {\bibfnamefont {A.}~\bibnamefont {Wallraff}}, \bibinfo {author}
  {\bibfnamefont {J.~M.}\ \bibnamefont {Gambetta}}, \bibinfo {author}
  {\bibfnamefont {A.}~\bibnamefont {Blais}}, \bibinfo {author} {\bibfnamefont
  {L.}~\bibnamefont {Frunzio}}, \bibinfo {author} {\bibfnamefont
  {J.}~\bibnamefont {Majer}}, \bibinfo {author} {\bibfnamefont
  {B.}~\bibnamefont {Johnson}}, \bibinfo {author} {\bibfnamefont {M.~H.}\
  \bibnamefont {Devoret}}, \bibinfo {author} {\bibfnamefont {S.~M.}\
  \bibnamefont {Girvin}}, \ and\ \bibinfo {author} {\bibfnamefont {R.~J.}\
  \bibnamefont {Schoelkopf}},\ }\href@noop {} {\bibfield  {journal} {\bibinfo
  {journal} {Nature}\ }\textbf {\bibinfo {volume} {445}},\ \bibinfo {pages}
  {515} (\bibinfo {year} {2007})}\BibitemShut {NoStop}%
\bibitem [{\citenamefont {Inomata}\ \emph {et~al.}(2016)\citenamefont
  {Inomata}, \citenamefont {Lin}, \citenamefont {Koshino}, \citenamefont
  {Oliver}, \citenamefont {Tsai}, \citenamefont {Yamamoto},\ and\ \citenamefont
  {Nakamura}}]{Inomata2016}%
  \BibitemOpen
  \bibfield  {author} {\bibinfo {author} {\bibfnamefont {K.}~\bibnamefont
  {Inomata}}, \bibinfo {author} {\bibfnamefont {Z.}~\bibnamefont {Lin}},
  \bibinfo {author} {\bibfnamefont {K.}~\bibnamefont {Koshino}}, \bibinfo
  {author} {\bibfnamefont {W.~D.}\ \bibnamefont {Oliver}}, \bibinfo {author}
  {\bibfnamefont {J.-S.}\ \bibnamefont {Tsai}}, \bibinfo {author}
  {\bibfnamefont {T.}~\bibnamefont {Yamamoto}}, \ and\ \bibinfo {author}
  {\bibfnamefont {Y.}~\bibnamefont {Nakamura}},\ }\href@noop {} {\bibfield
  {journal} {\bibinfo  {journal} {Nature Communications}\ }\textbf {\bibinfo
  {volume} {7}},\ \bibinfo {pages} {12303} (\bibinfo {year}
  {2016})}\BibitemShut {NoStop}%
\bibitem [{\citenamefont {Gu}\ \emph {et~al.}(2017)\citenamefont {Gu},
  \citenamefont {Kockum}, \citenamefont {Miranowicz}, \citenamefont {xi~Liu},\
  and\ \citenamefont {Nori}}]{GU20171}%
  \BibitemOpen
  \bibfield  {author} {\bibinfo {author} {\bibfnamefont {X.}~\bibnamefont
  {Gu}}, \bibinfo {author} {\bibfnamefont {A.~F.}\ \bibnamefont {Kockum}},
  \bibinfo {author} {\bibfnamefont {A.}~\bibnamefont {Miranowicz}}, \bibinfo
  {author} {\bibfnamefont {Y.}~\bibnamefont {xi~Liu}}, \ and\ \bibinfo {author}
  {\bibfnamefont {F.}~\bibnamefont {Nori}},\ }\href {\doibase
  https://doi.org/10.1016/j.physrep.2017.10.002} {\bibfield  {journal}
  {\bibinfo  {journal} {Physics Reports}\ }\textbf {\bibinfo {volume}
  {718-719}},\ \bibinfo {pages} {1 } (\bibinfo {year} {2017})},\ \bibinfo
  {note} {microwave photonics with superconducting quantum
  circuits}\BibitemShut {NoStop}%
\bibitem [{\citenamefont {Sathyamoorthy}\ \emph {et~al.}(2014)\citenamefont
  {Sathyamoorthy}, \citenamefont {Tornberg}, \citenamefont {Kockum},
  \citenamefont {Baragiola}, \citenamefont {Combes}, \citenamefont {Wilson},
  \citenamefont {Stace},\ and\ \citenamefont
  {Johansson}}]{PhysRevLett.112.093601}%
  \BibitemOpen
  \bibfield  {author} {\bibinfo {author} {\bibfnamefont {S.~R.}\ \bibnamefont
  {Sathyamoorthy}}, \bibinfo {author} {\bibfnamefont {L.}~\bibnamefont
  {Tornberg}}, \bibinfo {author} {\bibfnamefont {A.~F.}\ \bibnamefont
  {Kockum}}, \bibinfo {author} {\bibfnamefont {B.~Q.}\ \bibnamefont
  {Baragiola}}, \bibinfo {author} {\bibfnamefont {J.}~\bibnamefont {Combes}},
  \bibinfo {author} {\bibfnamefont {C.~M.}\ \bibnamefont {Wilson}}, \bibinfo
  {author} {\bibfnamefont {T.~M.}\ \bibnamefont {Stace}}, \ and\ \bibinfo
  {author} {\bibfnamefont {G.}~\bibnamefont {Johansson}},\ }\href {\doibase
  10.1103/PhysRevLett.112.093601} {\bibfield  {journal} {\bibinfo  {journal}
  {Phys. Rev. Lett.}\ }\textbf {\bibinfo {volume} {112}},\ \bibinfo {pages}
  {093601} (\bibinfo {year} {2014})}\BibitemShut {NoStop}%
\bibitem [{\citenamefont {Paik}\ \emph {et~al.}(2011)\citenamefont {Paik},
  \citenamefont {Schuster}, \citenamefont {Bishop}, \citenamefont {Kirchmair},
  \citenamefont {Catelani}, \citenamefont {Sears}, \citenamefont {Johnson},
  \citenamefont {Reagor}, \citenamefont {Frunzio}, \citenamefont {Glazman},
  \citenamefont {Girvin}, \citenamefont {Devoret},\ and\ \citenamefont
  {Schoelkopf}}]{PhysRevLett.107.240501}%
  \BibitemOpen
  \bibfield  {author} {\bibinfo {author} {\bibfnamefont {H.}~\bibnamefont
  {Paik}}, \bibinfo {author} {\bibfnamefont {D.~I.}\ \bibnamefont {Schuster}},
  \bibinfo {author} {\bibfnamefont {L.~S.}\ \bibnamefont {Bishop}}, \bibinfo
  {author} {\bibfnamefont {G.}~\bibnamefont {Kirchmair}}, \bibinfo {author}
  {\bibfnamefont {G.}~\bibnamefont {Catelani}}, \bibinfo {author}
  {\bibfnamefont {A.~P.}\ \bibnamefont {Sears}}, \bibinfo {author}
  {\bibfnamefont {B.~R.}\ \bibnamefont {Johnson}}, \bibinfo {author}
  {\bibfnamefont {M.~J.}\ \bibnamefont {Reagor}}, \bibinfo {author}
  {\bibfnamefont {L.}~\bibnamefont {Frunzio}}, \bibinfo {author} {\bibfnamefont
  {L.~I.}\ \bibnamefont {Glazman}}, \bibinfo {author} {\bibfnamefont {S.~M.}\
  \bibnamefont {Girvin}}, \bibinfo {author} {\bibfnamefont {M.~H.}\
  \bibnamefont {Devoret}}, \ and\ \bibinfo {author} {\bibfnamefont {R.~J.}\
  \bibnamefont {Schoelkopf}},\ }\href {\doibase 10.1103/PhysRevLett.107.240501}
  {\bibfield  {journal} {\bibinfo  {journal} {Phys. Rev. Lett.}\ }\textbf
  {\bibinfo {volume} {107}},\ \bibinfo {pages} {240501} (\bibinfo {year}
  {2011})}\BibitemShut {NoStop}%
\bibitem [{\citenamefont {Koch}\ \emph {et~al.}(2007)\citenamefont {Koch},
  \citenamefont {Yu}, \citenamefont {Gambetta}, \citenamefont {Houck},
  \citenamefont {Schuster}, \citenamefont {Majer}, \citenamefont {Blais},
  \citenamefont {Devoret}, \citenamefont {Girvin},\ and\ \citenamefont
  {Schoelkopf}}]{PhysRevA.76.042319}%
  \BibitemOpen
  \bibfield  {author} {\bibinfo {author} {\bibfnamefont {J.}~\bibnamefont
  {Koch}}, \bibinfo {author} {\bibfnamefont {T.~M.}\ \bibnamefont {Yu}},
  \bibinfo {author} {\bibfnamefont {J.}~\bibnamefont {Gambetta}}, \bibinfo
  {author} {\bibfnamefont {A.~A.}\ \bibnamefont {Houck}}, \bibinfo {author}
  {\bibfnamefont {D.~I.}\ \bibnamefont {Schuster}}, \bibinfo {author}
  {\bibfnamefont {J.}~\bibnamefont {Majer}}, \bibinfo {author} {\bibfnamefont
  {A.}~\bibnamefont {Blais}}, \bibinfo {author} {\bibfnamefont {M.~H.}\
  \bibnamefont {Devoret}}, \bibinfo {author} {\bibfnamefont {S.~M.}\
  \bibnamefont {Girvin}}, \ and\ \bibinfo {author} {\bibfnamefont {R.~J.}\
  \bibnamefont {Schoelkopf}},\ }\href {\doibase 10.1103/PhysRevA.76.042319}
  {\bibfield  {journal} {\bibinfo  {journal} {Phys. Rev. A}\ }\textbf {\bibinfo
  {volume} {76}},\ \bibinfo {pages} {042319} (\bibinfo {year}
  {2007})}\BibitemShut {NoStop}%
\bibitem [{\citenamefont {Wendin}(2017)}]{wendin2017quantum}%
  \BibitemOpen
  \bibfield  {author} {\bibinfo {author} {\bibfnamefont {G.}~\bibnamefont
  {Wendin}},\ }\href@noop {} {\bibfield  {journal} {\bibinfo  {journal}
  {Reports on Progress in Physics}\ }\textbf {\bibinfo {volume} {80}},\
  \bibinfo {pages} {106001} (\bibinfo {year} {2017})}\BibitemShut {NoStop}%
\bibitem [{\citenamefont {Wallraff}\ \emph {et~al.}(2004)\citenamefont
  {Wallraff}, \citenamefont {Schuster}, \citenamefont {Blais}, \citenamefont
  {Frunzio}, \citenamefont {Huang}, \citenamefont {Majer}, \citenamefont
  {Kumar}, \citenamefont {Girvin},\ and\ \citenamefont
  {Schoelkopf}}]{wallraff2004strong}%
  \BibitemOpen
  \bibfield  {author} {\bibinfo {author} {\bibfnamefont {A.}~\bibnamefont
  {Wallraff}}, \bibinfo {author} {\bibfnamefont {D.~I.}\ \bibnamefont
  {Schuster}}, \bibinfo {author} {\bibfnamefont {A.}~\bibnamefont {Blais}},
  \bibinfo {author} {\bibfnamefont {L.}~\bibnamefont {Frunzio}}, \bibinfo
  {author} {\bibfnamefont {R.-S.}\ \bibnamefont {Huang}}, \bibinfo {author}
  {\bibfnamefont {J.}~\bibnamefont {Majer}}, \bibinfo {author} {\bibfnamefont
  {S.}~\bibnamefont {Kumar}}, \bibinfo {author} {\bibfnamefont {S.~M.}\
  \bibnamefont {Girvin}}, \ and\ \bibinfo {author} {\bibfnamefont {R.~J.}\
  \bibnamefont {Schoelkopf}},\ }\href@noop {} {\bibfield  {journal} {\bibinfo
  {journal} {Nature}\ }\textbf {\bibinfo {volume} {431}},\ \bibinfo {pages}
  {162} (\bibinfo {year} {2004})}\BibitemShut {NoStop}%
\bibitem [{\citenamefont {Il'ichev}\ \emph {et~al.}(2003)\citenamefont
  {Il'ichev}, \citenamefont {Oukhanski}, \citenamefont {Izmalkov},
  \citenamefont {Wagner}, \citenamefont {Grajcar}, \citenamefont {Meyer},
  \citenamefont {Smirnov}, \citenamefont {Maassen van~den Brink}, \citenamefont
  {Amin},\ and\ \citenamefont {Zagoskin}}]{PhysRevLett.91.097906}%
  \BibitemOpen
  \bibfield  {author} {\bibinfo {author} {\bibfnamefont {E.}~\bibnamefont
  {Il'ichev}}, \bibinfo {author} {\bibfnamefont {N.}~\bibnamefont {Oukhanski}},
  \bibinfo {author} {\bibfnamefont {A.}~\bibnamefont {Izmalkov}}, \bibinfo
  {author} {\bibfnamefont {T.}~\bibnamefont {Wagner}}, \bibinfo {author}
  {\bibfnamefont {M.}~\bibnamefont {Grajcar}}, \bibinfo {author} {\bibfnamefont
  {H.-G.}\ \bibnamefont {Meyer}}, \bibinfo {author} {\bibfnamefont {A.~Y.}\
  \bibnamefont {Smirnov}}, \bibinfo {author} {\bibfnamefont {A.}~\bibnamefont
  {Maassen van~den Brink}}, \bibinfo {author} {\bibfnamefont {M.~H.~S.}\
  \bibnamefont {Amin}}, \ and\ \bibinfo {author} {\bibfnamefont {A.~M.}\
  \bibnamefont {Zagoskin}},\ }\href {\doibase 10.1103/PhysRevLett.91.097906}
  {\bibfield  {journal} {\bibinfo  {journal} {Phys. Rev. Lett.}\ }\textbf
  {\bibinfo {volume} {91}},\ \bibinfo {pages} {097906} (\bibinfo {year}
  {2003})}\BibitemShut {NoStop}%
\bibitem [{\citenamefont {Hoi}\ \emph {et~al.}(2013)\citenamefont {Hoi},
  \citenamefont {Wilson}, \citenamefont {Johansson}, \citenamefont {Lindkvist},
  \citenamefont {Peropadre}, \citenamefont {Palomaki},\ and\ \citenamefont
  {Delsing}}]{Hoi_2013}%
  \BibitemOpen
  \bibfield  {author} {\bibinfo {author} {\bibfnamefont {I.-C.}\ \bibnamefont
  {Hoi}}, \bibinfo {author} {\bibfnamefont {C.~M.}\ \bibnamefont {Wilson}},
  \bibinfo {author} {\bibfnamefont {G.}~\bibnamefont {Johansson}}, \bibinfo
  {author} {\bibfnamefont {J.}~\bibnamefont {Lindkvist}}, \bibinfo {author}
  {\bibfnamefont {B.}~\bibnamefont {Peropadre}}, \bibinfo {author}
  {\bibfnamefont {T.}~\bibnamefont {Palomaki}}, \ and\ \bibinfo {author}
  {\bibfnamefont {P.}~\bibnamefont {Delsing}},\ }\href {\doibase
  10.1088/1367-2630/15/2/025011} {\bibfield  {journal} {\bibinfo  {journal}
  {New Journal of Physics}\ }\textbf {\bibinfo {volume} {15}},\ \bibinfo
  {pages} {025011} (\bibinfo {year} {2013})}\BibitemShut {NoStop}%
\bibitem [{\citenamefont {Rakhmanov}\ \emph {et~al.}(2008)\citenamefont
  {Rakhmanov}, \citenamefont {Zagoskin}, \citenamefont {Savel'ev},\ and\
  \citenamefont {Nori}}]{rakhmanov2008quantum}%
  \BibitemOpen
  \bibfield  {author} {\bibinfo {author} {\bibfnamefont {A.~L.}\ \bibnamefont
  {Rakhmanov}}, \bibinfo {author} {\bibfnamefont {A.~M.}\ \bibnamefont
  {Zagoskin}}, \bibinfo {author} {\bibfnamefont {S.}~\bibnamefont {Savel'ev}},
  \ and\ \bibinfo {author} {\bibfnamefont {F.}~\bibnamefont {Nori}},\
  }\href@noop {} {\bibfield  {journal} {\bibinfo  {journal} {Physical Review
  B}\ }\textbf {\bibinfo {volume} {77}},\ \bibinfo {pages} {144507} (\bibinfo
  {year} {2008})}\BibitemShut {NoStop}%
\bibitem [{\citenamefont {Zagoskin}\ \emph {et~al.}(2009)\citenamefont
  {Zagoskin}, \citenamefont {Rakhmanov}, \citenamefont {Savel'ev},\ and\
  \citenamefont {Nori}}]{zagoskin2009quantum}%
  \BibitemOpen
  \bibfield  {author} {\bibinfo {author} {\bibfnamefont {A.}~\bibnamefont
  {Zagoskin}}, \bibinfo {author} {\bibfnamefont {A.}~\bibnamefont {Rakhmanov}},
  \bibinfo {author} {\bibfnamefont {S.}~\bibnamefont {Savel'ev}}, \ and\
  \bibinfo {author} {\bibfnamefont {F.}~\bibnamefont {Nori}},\ }\href@noop {}
  {\bibfield  {journal} {\bibinfo  {journal} {physica status solidi (b)}\
  }\textbf {\bibinfo {volume} {246}},\ \bibinfo {pages} {955} (\bibinfo {year}
  {2009})}\BibitemShut {NoStop}%
\bibitem [{\citenamefont {Savel'ev}\ \emph {et~al.}(2012)\citenamefont
  {Savel'ev}, \citenamefont {Zagoskin}, \citenamefont {Rakhmanov},
  \citenamefont {Omelyanchouk}, \citenamefont {Washington},\ and\ \citenamefont
  {Nori}}]{savel2012two}%
  \BibitemOpen
  \bibfield  {author} {\bibinfo {author} {\bibfnamefont {S.}~\bibnamefont
  {Savel'ev}}, \bibinfo {author} {\bibfnamefont {A.}~\bibnamefont {Zagoskin}},
  \bibinfo {author} {\bibfnamefont {A.}~\bibnamefont {Rakhmanov}}, \bibinfo
  {author} {\bibfnamefont {A.}~\bibnamefont {Omelyanchouk}}, \bibinfo {author}
  {\bibfnamefont {Z.}~\bibnamefont {Washington}}, \ and\ \bibinfo {author}
  {\bibfnamefont {F.}~\bibnamefont {Nori}},\ }\href@noop {} {\bibfield
  {journal} {\bibinfo  {journal} {Physical Review A}\ }\textbf {\bibinfo
  {volume} {85}},\ \bibinfo {pages} {013811} (\bibinfo {year}
  {2012})}\BibitemShut {NoStop}%
\bibitem [{\citenamefont {Ivi{\'c}}\ \emph {et~al.}(2016)\citenamefont
  {Ivi{\'c}}, \citenamefont {Lazarides},\ and\ \citenamefont
  {Tsironis}}]{ivic2016qubit}%
  \BibitemOpen
  \bibfield  {author} {\bibinfo {author} {\bibfnamefont {Z.}~\bibnamefont
  {Ivi{\'c}}}, \bibinfo {author} {\bibfnamefont {N.}~\bibnamefont {Lazarides}},
  \ and\ \bibinfo {author} {\bibfnamefont {G.}~\bibnamefont {Tsironis}},\
  }\href@noop {} {\bibfield  {journal} {\bibinfo  {journal} {Scientific
  reports}\ }\textbf {\bibinfo {volume} {6}},\ \bibinfo {pages} {29374}
  (\bibinfo {year} {2016})}\BibitemShut {NoStop}%
\bibitem [{\citenamefont {Grimsmo}\ \emph {et~al.}(2020)\citenamefont
  {Grimsmo}, \citenamefont {Royer}, \citenamefont {Kreikebaum}, \citenamefont
  {Ye}, \citenamefont {O'Brien}, \citenamefont {Siddiqi},\ and\ \citenamefont
  {Blais}}]{grimsmo2020quantum}%
  \BibitemOpen
  \bibfield  {author} {\bibinfo {author} {\bibfnamefont {A.~L.}\ \bibnamefont
  {Grimsmo}}, \bibinfo {author} {\bibfnamefont {B.}~\bibnamefont {Royer}},
  \bibinfo {author} {\bibfnamefont {J.~M.}\ \bibnamefont {Kreikebaum}},
  \bibinfo {author} {\bibfnamefont {Y.}~\bibnamefont {Ye}}, \bibinfo {author}
  {\bibfnamefont {K.}~\bibnamefont {O'Brien}}, \bibinfo {author} {\bibfnamefont
  {I.}~\bibnamefont {Siddiqi}}, \ and\ \bibinfo {author} {\bibfnamefont
  {A.}~\bibnamefont {Blais}},\ }\href@noop {} {\enquote {\bibinfo {title}
  {Quantum metamaterial for nondestructive microwave photon counting},}\ }
  (\bibinfo {year} {2020}),\ \Eprint {http://arxiv.org/abs/2005.06483}
  {arXiv:2005.06483 [quant-ph]} \BibitemShut {NoStop}%
\bibitem [{\citenamefont {Zagoskin}\ \emph {et~al.}(2013)\citenamefont
  {Zagoskin}, \citenamefont {Wilson}, \citenamefont {Everitt}, \citenamefont
  {Savel'ev}, \citenamefont {Gulevich}, \citenamefont {Allen}, \citenamefont
  {Dubrovich},\ and\ \citenamefont {Il'ichev}}]{Zagoskin_2013}%
  \BibitemOpen
  \bibfield  {author} {\bibinfo {author} {\bibfnamefont {A.~M.}\ \bibnamefont
  {Zagoskin}}, \bibinfo {author} {\bibfnamefont {R.~D.}\ \bibnamefont
  {Wilson}}, \bibinfo {author} {\bibfnamefont {M.}~\bibnamefont {Everitt}},
  \bibinfo {author} {\bibfnamefont {S.}~\bibnamefont {Savel'ev}}, \bibinfo
  {author} {\bibfnamefont {D.~R.}\ \bibnamefont {Gulevich}}, \bibinfo {author}
  {\bibfnamefont {J.}~\bibnamefont {Allen}}, \bibinfo {author} {\bibfnamefont
  {V.~K.}\ \bibnamefont {Dubrovich}}, \ and\ \bibinfo {author} {\bibfnamefont
  {E.}~\bibnamefont {Il'ichev}},\ }\href {\doibase 10.1038/srep03464}
  {\bibfield  {journal} {\bibinfo  {journal} {Scientific Reports}\ }\textbf
  {\bibinfo {volume} {3}} (\bibinfo {year} {2013}),\
  10.1038/srep03464}\BibitemShut {NoStop}%
\bibitem [{\citenamefont {Giovannetti}\ \emph {et~al.}(2006)\citenamefont
  {Giovannetti}, \citenamefont {Lloyd},\ and\ \citenamefont
  {Maccone}}]{Giovannetti2006}%
  \BibitemOpen
  \bibfield  {author} {\bibinfo {author} {\bibfnamefont {V.}~\bibnamefont
  {Giovannetti}}, \bibinfo {author} {\bibfnamefont {S.}~\bibnamefont {Lloyd}},
  \ and\ \bibinfo {author} {\bibfnamefont {L.}~\bibnamefont {Maccone}},\
  }\href@noop {} {\bibfield  {journal} {\bibinfo  {journal} {Phys. Rev. Lett.}\
  }\textbf {\bibinfo {volume} {96}},\ \bibinfo {pages} {010401} (\bibinfo
  {year} {2006})}\BibitemShut {NoStop}%
\bibitem [{\citenamefont {Giovannetti}\ \emph {et~al.}(2004)\citenamefont
  {Giovannetti}, \citenamefont {Lloyd},\ and\ \citenamefont
  {Maccone}}]{Giovannetti2004}%
  \BibitemOpen
  \bibfield  {author} {\bibinfo {author} {\bibfnamefont {V.}~\bibnamefont
  {Giovannetti}}, \bibinfo {author} {\bibfnamefont {S.}~\bibnamefont {Lloyd}},
  \ and\ \bibinfo {author} {\bibfnamefont {L.}~\bibnamefont {Maccone}},\
  }\href@noop {} {\bibfield  {journal} {\bibinfo  {journal} {Science}\ }\textbf
  {\bibinfo {volume} {306}},\ \bibinfo {pages} {1330} (\bibinfo {year}
  {2004})}\BibitemShut {NoStop}%
\bibitem [{\citenamefont {Shlyakhov}\ \emph {et~al.}(2018)\citenamefont
  {Shlyakhov}, \citenamefont {Zemlyanov}, \citenamefont {Suslov}, \citenamefont
  {Lebedev}, \citenamefont {Paraoanu}, \citenamefont {Lesovik},\ and\
  \citenamefont {Blatter}}]{PhysRevA.97.022115}%
  \BibitemOpen
  \bibfield  {author} {\bibinfo {author} {\bibfnamefont {A.~R.}\ \bibnamefont
  {Shlyakhov}}, \bibinfo {author} {\bibfnamefont {V.~V.}\ \bibnamefont
  {Zemlyanov}}, \bibinfo {author} {\bibfnamefont {M.~V.}\ \bibnamefont
  {Suslov}}, \bibinfo {author} {\bibfnamefont {A.~V.}\ \bibnamefont {Lebedev}},
  \bibinfo {author} {\bibfnamefont {G.~S.}\ \bibnamefont {Paraoanu}}, \bibinfo
  {author} {\bibfnamefont {G.~B.}\ \bibnamefont {Lesovik}}, \ and\ \bibinfo
  {author} {\bibfnamefont {G.}~\bibnamefont {Blatter}},\ }\href {\doibase
  10.1103/PhysRevA.97.022115} {\bibfield  {journal} {\bibinfo  {journal} {Phys.
  Rev. A}\ }\textbf {\bibinfo {volume} {97}},\ \bibinfo {pages} {022115}
  (\bibinfo {year} {2018})}\BibitemShut {NoStop}%
\bibitem [{\citenamefont {Danilin}\ \emph {et~al.}(2018)\citenamefont
  {Danilin}, \citenamefont {Lebedev}, \citenamefont {Vepsäläinen},
  \citenamefont {Lesovik}, \citenamefont {Blatter},\ and\ \citenamefont
  {Paraoanu}}]{Danilin_2018}%
  \BibitemOpen
  \bibfield  {author} {\bibinfo {author} {\bibfnamefont {S.}~\bibnamefont
  {Danilin}}, \bibinfo {author} {\bibfnamefont {A.~V.}\ \bibnamefont
  {Lebedev}}, \bibinfo {author} {\bibfnamefont {A.}~\bibnamefont
  {Vepsäläinen}}, \bibinfo {author} {\bibfnamefont {G.~B.}\ \bibnamefont
  {Lesovik}}, \bibinfo {author} {\bibfnamefont {G.}~\bibnamefont {Blatter}}, \
  and\ \bibinfo {author} {\bibfnamefont {G.~S.}\ \bibnamefont {Paraoanu}},\
  }\href {\doibase 10.1038/s41534-018-0078-y} {\bibfield  {journal} {\bibinfo
  {journal} {npj Quantum Information}\ }\textbf {\bibinfo {volume} {4}}
  (\bibinfo {year} {2018}),\ 10.1038/s41534-018-0078-y}\BibitemShut {NoStop}%
\bibitem [{\citenamefont {Sørensen}\ \emph {et~al.}(2001)\citenamefont
  {Sørensen}, \citenamefont {Duan}, \citenamefont {Cirac},\ and\ \citenamefont
  {Zoller}}]{Cirac}%
  \BibitemOpen
  \bibfield  {author} {\bibinfo {author} {\bibfnamefont {A.}~\bibnamefont
  {Sørensen}}, \bibinfo {author} {\bibfnamefont {L.-M.}\ \bibnamefont {Duan}},
  \bibinfo {author} {\bibfnamefont {J.~I.}\ \bibnamefont {Cirac}}, \ and\
  \bibinfo {author} {\bibfnamefont {P.}~\bibnamefont {Zoller}},\ }\href
  {\doibase 10.1038/35051038} {\bibfield  {journal} {\bibinfo  {journal}
  {Nature}\ }\textbf {\bibinfo {volume} {409}},\ \bibinfo {pages} {63}
  (\bibinfo {year} {2001})}\BibitemShut {NoStop}%
\bibitem [{\citenamefont {Andr\'e}\ and\ \citenamefont
  {Lukin}(2002)}]{PhysRevA.65.053819}%
  \BibitemOpen
  \bibfield  {author} {\bibinfo {author} {\bibfnamefont {A.}~\bibnamefont
  {Andr\'e}}\ and\ \bibinfo {author} {\bibfnamefont {M.~D.}\ \bibnamefont
  {Lukin}},\ }\href {\doibase 10.1103/PhysRevA.65.053819} {\bibfield  {journal}
  {\bibinfo  {journal} {Phys. Rev. A}\ }\textbf {\bibinfo {volume} {65}},\
  \bibinfo {pages} {053819} (\bibinfo {year} {2002})}\BibitemShut {NoStop}%
\end{thebibliography}%
%\bibliography{REFERENCES2019.bib}

\end{document}